\documentclass[twocolumn,showpacs,prd,floatfix,axodraw]{revtex4}
\usepackage{mathrsfs}
\usepackage{graphicx,,booktabs,bm}
\usepackage{overpic}
\usepackage{color}
\usepackage{amssymb}

\usepackage{mathrsfs,bm,amsmath,amssymb}
\usepackage{longtable,lscape}
\usepackage{txfonts}
\usepackage{amssymb}
\usepackage{indentfirst}
\usepackage{graphicx,,booktabs}
\usepackage{multirow,ulem}

\usepackage{color}
\usepackage{amssymb}

\definecolor{cover}{rgb}{0.77,0.87,0.88}
\definecolor{blueone}{rgb}{0.1,0.1,.7}
\definecolor{citec}{rgb}{0.14,0.47,0.09}
\definecolor{two}{rgb}{0.0,0.5,0.}
\definecolor{three}{rgb}{.5,.1,0.15}
\usepackage[bookmarks=true,bookmarksopen=false,plainpages=false,breaklinks=true,
   bookmarksnumbered=true,hypertexnames=false,
   filecolor=blue,urlcolor=three,menucolor=three,
   linkcolor=three,citecolor=blueone, colorlinks,
   anchorcolor=blue,runcolor=pink,frenchlinks=red
   pdfstartview=FitH,pdftitle=title,%
   pdfauthor=author]{hyperref}

\begin{document}
\title{Investigating the $P$-wave $DK^{*}$  Molecular Interpretation of the $D_{s1}(2700)$ via Its Strong Decays}
\author{Qing Lu}
\affiliation{Key Laboratory of Computational Physics of Sichuan Province, School of Mathematics and Physics, Yibin University, Yibin 644000, China}
\author{ZheHao Zhu}
\affiliation{School of Physical Science and Technology, Southwest Jiaotong University, Chengdu 610031,China}
\author{Cai Cheng}
\affiliation{School of Physics and Electronic Engineering, Sichuan Normal University, Chengdu 610101, China}
\author{Yin Huang\footnote{corresponding author}} \email{huangy2019@swjtu.edu.cn}
\affiliation{School of Physical Science and Technology, Southwest Jiaotong University, Chengdu 610031,China}

\begin{abstract}
In this work, we investigate the possibility of interpreting the $D_{s1}(2700)$ as a $P$-wave $DK^{*}$ hadronic molecular state through a systematic study of its strong decay properties.
Using the effective Lagrangian approach, we calculate the strong decay widths of the $D_{s1}(2700)$ into the two-body final states $DK$, $D^{*}K$, $D_s\eta$, and $D_s^{*}\eta$, as well as
the three-body $D\pi K$ channel. The coupling of the $D_{s1}(2700)$ to its constituents, $DK^{*}$, is constrained by the available experimental measurement of $R_{D_{s1}} =\Gamma(D_{s1}(2700)
\to D^{*}K)/\Gamma(D_{s1}(2700)\to DK)=0.91\pm0.13\pm0.12$. With the resulting coupling, the total decay width can be readily obtained and is found to be in good agreement with the experimental
measurement, providing strong support for interpreting the $D_{s1}(2700)$ as a $P$-wave $DK^{*}$ molecular state. The experimentally unobserved $D_{s1}(2700)\to D_s^{*}\eta$ decay channel
provides a further test of this interpretation, as its predicted sizable decay width differs significantly from the conventional quark model prediction.
\end{abstract}

\date{\today}


\maketitle
\section{Introduction}\label{sec:intro}
The hadronic molecular picture, exemplified by the deuteron--a loosely bound proton-neutron system--has emerged as an important framework for understanding the
internal structures and formation mechanisms of exotic hadrons.  Exotic states refer to hadrons that cannot be accommodated within the conventional quark model,
where mesons are composed of quark--antiquark pairs and baryons consist of three quarks. The discovery of the $X(3872)$ in 2003~\cite{Belle:2003nnu}
marked a milestone in this field, and it is now widely regarded as a promising $D\bar{D}^{*}$ molecular candidate~\cite{Brambilla:2019esw,Chen:2022asf,Meng:2022ozq}.
Following this discovery, several hidden-charm pentaquark candidates, including the $P_c$ and $P_{cs}$ states observed by the LHCb Collaboration~\cite{LHCb:2015yax,LHCb:2016ztz,LHCb:2016lve,LHCb:2019kea,LHCb:2020jpq,LHCb:2022ogu}, have also been proposed as hadronic molecules,
with possible configurations such as $D^{(*)}\Sigma_c^{(*)}$ and $D^{(*)}\Xi_c$~\cite{Chen:2019bip,Guo:2019fdo,Xiao:2019aya,He:2019ify,Xiao:2019mvs,
Roca:2015dva,Chen:2015moa,Chen:2015loa,Yang:2015bmv,Huang:2015uda,Du:2019pij}.  These molecular candidates are generally interpreted as $S$-wave bound states,
where the absence of a centrifugal barrier associated with the relative orbital angular momentum provides the most favorable condition for molecular binding.
In contrast, $P$-wave and higher partial-wave hadronic molecules remain largely unexplored. However, the existence of a non-negligible $D$-wave component in the
deuteron~\cite{Machleidt:2000ge}, the archetypal hadronic molecule, demonstrates that non-$S$-wave configurations can also exist in nature.

In 2004, the $G(3900)$ structure was experimentally observed~\cite{BESIII:2024ths} and has been proposed as a possible $P$-wave $D\bar{D}^{*}/D^{*}\bar{D}$ molecular
candidate~\cite{Lin:2024qcq}.  The experimentally observed $\Xi(2030)$ has also been proposed as a possible $P$-wave hadronic molecular state~\cite{Hei:2023eqz,Feng:2024jzu},
dominantly composed of a $\bar{K}^{*}\Sigma$ configuration. The charmed-strange sector may provide promising candidates for $P$-wave molecular states.
To date, eleven charmed-strange meson states have been experimentally established~\cite{ParticleDataGroup:2024cfk}.
Among them, the ground-state pseudoscalar meson $D_s(1968)$ and its vector partner $D_s^*(2112)$ are well described
within the conventional quark model as $c\bar{s}$ states~\cite{Godfrey:1985xj}, corresponding to the $1^1S_0$ and $1^3S_1$
configurations, respectively.  However, some of the remaining charmed-strange mesons, including $D_{s0}^*(2317)$, $D_{s1}(2460)$,
$D_{s1}(2536)$, $D_{s2}^*(2573)$, $D_{s0}(2590)$, $D_{s1}(2700)$, $D_{s1/s3}(2860)$, and $D_{sJ}(3040)$, display unexpected
features that are difficult to reconcile with a simple
quark-antiquark interpretation.  In particular, the low-lying $D_{s0}^*(2317)$ and $D_{s1}(2460)$ states have attracted considerable attention,
since their masses are found to be approximately 160 MeV and 70 MeV lower, respectively, than the corresponding predictions from the conventional
quark model~\cite{Godfrey:1985xj}. Due to their masses being remarkably close to the $DK$ and $D^*K$ thresholds, respectively, these states are
expected to contain significant hadronic molecular components.  Consequently, extensive theoretical investigations have
explored the possibility that $D_{s0}^*(2317)$ and $D_{s1}(2460)$ can be interpreted as $DK$ and $D^*K$ molecular states, respectively~\cite{BaBar:2003oey,CLEO:2003ggt,Belle:2003kup,Belle:2003guh,BaBar:2004yux,Meng:2022ozq,Xie:2010zza,Guo:2006fu,Guo:2006rp,Gamermann:2006nm,
Zhu:2019vnr,Mohler:2013rwa,Altenbuchinger:2013vwa,Faessler:2007gv,Faessler:2007us,Cleven:2014oka,Xiao:2016hoa}.
More importantly, the $DK$ and $D^*K$ molecular interpretations naturally account for the observed isospin-violating decay modes of $D_{s0}^*(2317)$
and $D_{s1}(2460)$ into $D_s^{(*)}\pi$~\cite{Xie:2010zza,Faessler:2007gv,Faessler:2007us,Cleven:2014oka,Xiao:2016hoa}.

Furthermore, beyond the well-known $D_{s0}^*(2317)$ and $D_{s1}(2460)$ molecular candidates, several other excited charmed-strange states have also been
suggested to possess molecular components. The $D_{s1}(2700)$ state is one of these candidates. The $D_{s1}(2700)$ state, with a mass of $2714\pm5$ MeV and
a total width of $122\pm10$ MeV according to the Particle Data Group (PDG)~\cite{ParticleDataGroup:2024cfk}, has been proposed as a molecular state dominated
by the $DK^{*}$ component in Ref.~\cite{Hao:2022vwt}. However, this molecular interpretation has received relatively limited attention.  Instead, the $D_{s1}(2700)$
state is generally interpreted as a conventional $c\bar{s}$ meson, mainly because quark-model calculation, although involving different quark structures,
can naturally reproduce the experimentally measured ratio of the partial decay widths~\cite{ParticleDataGroup:2024cfk},
\[
R_{D_{s1}}=\frac{\Gamma(D_{s1}(2700)\to D^{*}K)}
{\Gamma(D_{s1}(2700)\to DK)}
=0.91\pm0.13\pm0.12 .
\]
For example, within the leading-order (LO) heavy hadron chiral perturbation theory (HHChPT), a model-independent approach, the decay properties of
the $D_{s1}(2700)$ under the $(2S,J^P=1^-)$ assignment were studied.  The predicted decay width ratio, $R_{D_{s1}(2700)}=0.91\pm0.04$~\cite{Colangelo:2007ds},
agrees remarkably well with the experimental result. Moreover, the observed mass of the $D_{s1}(2700)$ is also in good agreement with the prediction
of the Godfrey--Isgur quark model~\cite{Godfrey:1985xj}. These findings provide strong support for interpreting the $D_{s1}(2700)$ as a conventional
$(2S,1^-)$ $c\bar{s}$ meson.

However, alternative conventional $c\bar{s}$ interpretations have also been proposed, in which the $D_{s1}(2700)$ is assigned as a $1^-(1\,{}^3D_1)$ state.
This assignment was supported by the $^3P_0$ model calculations of its decay properties and the comparison with available experimental data~\cite{Zhang:2006yj}.
 Nevertheless, the possibility that the $D_{s1}(2700)$ corresponds to the conventional $1^-(2\,{}^3S_1)$ $c\bar{s}$ configuration was not excluded.
 By investigating its spectrum and decay properties within a new theoretical framework, the authors of Ref.~\cite{Close:2006gr} proposed that the $D_{s1}(2700)$
 could be interpreted as a mixed state involving the $(2S,\,{}^3S_1)$ and $(1D,\,{}^3D_1)$ $c\bar{s}$ configurations. In this scenario, the $B^0\to D_{s1}^{+}D^{(*)-}$
 decay modes were predicted to have sizable branching fractions, suggesting their potential for experimentally searching for the $D_{s1}(2700)$. This mixing
 interpretation was further supported by Refs.~\cite{Zhong:2009sk,Li:2009qu,Song:2015nia}, where the decay properties of the $D_{s1}(2700)$ were investigated
 within different theoretical models, and the theoretical predictions were found to be consistent with the available experimental data.

The diversity of these conventional $c\bar{s}$ interpretations suggests that the nature of the $D_{s1}(2700)$ deserves further investigation. In particular,
its mass lies close to the $DK^{*}$ threshold, which motivates us to explore the possibility of a $DK^{*}$ molecular interpretation. By studying the $DK^{*}$
interaction, we find that the $D_{s1}(2700)$ can be accommodated as a $DK^{*}$ molecular state~\cite{Jiang:2026rpu}. With the spin-parity assignment $J^P=1^-$,
this state corresponds to a $P$-wave $DK^{*}$ molecule.  However, the mass spectrum alone cannot provide a definitive identification of a molecular state, and
its decay properties are crucial for revealing the underlying structure. In this work, we study the strong decay properties of the $D_{s1}(2700)$ assuming a
$DK^{*}$ molecular configuration, aiming to examine the viability of this interpretation. The calculations are performed within the hadronic molecule framework
using effective Lagrangians.

This paper is organized as follows. In Sec.~\ref{Sec: formulism},  we will present the theoretical  formalism.  In Sec.~\ref{Sec: results}, the numerical result
will be given, followed by discussions and conclusions in the last section.
\section{FORMALISM AND INGREDIENTS}\label{Sec: formulism}
\begin{figure}[http]
\begin{center}
\includegraphics[bb=80 350 1050 710, clip, scale=0.55]{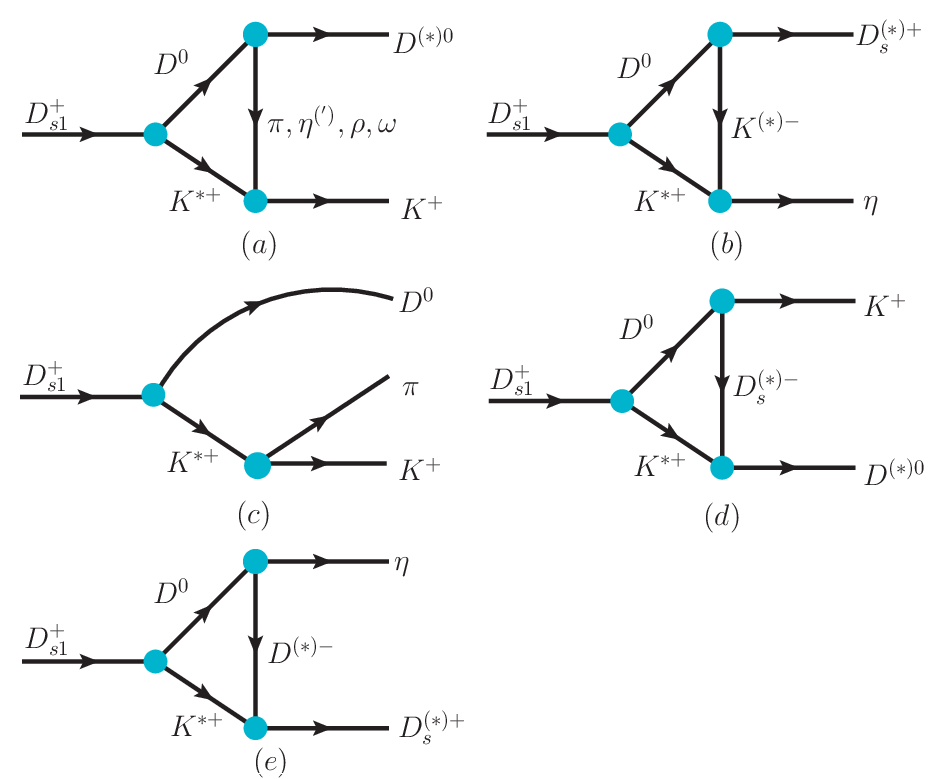}
\caption{The relevant two-body and three-body decay diagrams of the $D_{s1}(2700)$ state within $DK^{*}$ molecular configuration. For the two-body decay modes,
we consider the contributions from the $t$-channel exchange of pseudoscalar mesons ($\pi$, $\eta$, $K$, $D$, and $D_s$) and vector mesons ($\rho$, $\omega$,
$\phi$, $D^{*}$ and $D_s^{*}$).}\label{cc1}
\end{center}
\end{figure}

In this work, we study the strong decays of the $P$-wave $DK^{*}$ molecular state to investigate whether the experimentally observed $D_{s1}(2700)$ state can be
interpreted as $DK^{*}$ molecular state. We focus on the experimentally observed two-body decay modes, $DK$ and $D^{*}K$, whose corresponding Feynman diagrams are
shown in Fig.~\ref{cc1}. Specifically, Fig.~\ref{cc1}(a) corresponds to the decays into the $D^{(*)0}K^{+}$ final states via the $t$-channel exchange of pseudoscalar
mesons ($\pi$, $\eta$, and $\eta'$) and vector mesons ($\rho$ and $\omega$), whereas Fig.~\ref{cc1}(d) represents the decays into the $K^{+}D^{0}$ or $K^{+}D^{*0}$
final states through the $t$-channel exchange of $D_s$ or $D_s^{*}$ mesons.  In addition, we investigate several decay modes that have not yet been observed experimentally
but are expected within the $DK^{*}$ molecular interpretation of the $D_{s1}(2700)$ state.  These include the $\eta D^{(*)}_s$, $\omega D^{(*)}_s$ (see Fig.~\ref{cc1}
(b and e)) decay channels, and the three-body $K\pi D^{0}$ decay modes (see Fig.~\ref{cc1} (c)).  Since the width of the $D^{*}$ meson is much smaller than that of the
$K^{*}$ meson, we treat the $D^{*}$ meson as a stable particle in the calculations of the three-body decay processes.

To calculate the decay widths corresponding to the diagrams shown in Fig.~\ref{cc1}, we employ the effective Lagrangian approach. Throughout this work, the
$D_{s1}(2700)$ state is denoted as $D_{s1}$ and interpreted as a $P$-wave $DK^{*}$ molecular state.   Accordingly, the simplest effective Lagrangian densities
for the $D_{s1}DK^{*}$ interaction can be constructed as~\cite{Fajfer:1992hi}.
\begin{align}
{\cal{L}}_{D_{s1}}&= g_{D_{s1}}\epsilon_{\mu\nu\alpha\beta}\partial^{\mu}D^{\nu}_{s1}(x)\int d^4y \Phi(y^2)\nonumber\\
                        &\times{}D(x+\omega_{K^{*}}y)\partial^{\alpha}K^{*\beta}(x-\omega_D y)\label{eq1},
\end{align}
where $\omega_{ij}=m_i/(m_i+m_j)$, with $m_i$ representing the masses of the $D$ and $K^{*}$ mesons.  The correlation function $\Phi(y^2)$ is
introduced to describe the distribution of the $D$ and $K^{*}$ components inside the hadronic molecular $D_{s1}$ state. Moreover,
it plays an essential role in regularizing the ultraviolet divergences arising from the triangle Feynman diagrams depicted in Fig.~\ref{cc1}. As the
relative coordinate $y$ increases, the constituent components of the molecular state become spatially separated, leading to the gradual disappearance
of the molecular structure. This behavior should be reflected in the amplitudes corresponding to the Feynman diagrams shown in Fig.~\ref{cc1}, which
are expected to vanish in the limit of $y\rightarrow\infty$.  Therefore, the correlation function $\Phi(y^2)$ is usually parametrized in the following
Gaussian form:
\begin{equation}
\Phi(p^2)\doteq\exp(-p_E^2/\Lambda^2)\label{eq2},
\end{equation}
where $p_E$ denotes the Euclidean Jacobi momentum. The parameter $\Lambda$ is the size parameter that characterizes the spatial distribution of the
constituent components inside the molecular state. Since $\Lambda$ cannot be determined from first principles, it is treated as a phenomenological
parameter and will be discussed later.

\begin{figure}[http]
\begin{center}
\includegraphics[bb=80 605 1050 710, clip, scale=0.58]{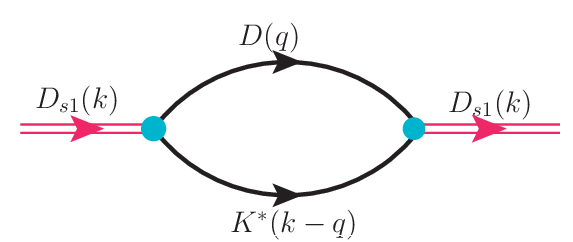}
\caption{Self-energy of $D_{s1}$ state. }
\label{cc2}
\end{center}
\end{figure}

The coupling constants $g_{D_{s1}}$  can be determined by the Weinberg compositeness condition~\cite{Weinberg:1962hj,Salam:1962ap},
which implies that the wave function renormalization constant of $D_{s1}$ is zero,
\begin{align}
Z_{D_{s1}}=1-\frac{d\Sigma^{T}_{D_{s1}}(k^2)}{dk^2}|_{k^2=m^2_{D_{s1}}}=0\label{eq3},
\end{align}
where $k$ is the four-momentum of the $D_{s1}$ states, and $m_{D_{s1}}$ denotes the mass of the corresponding molecular state. $\Sigma^{T}$ denotes
the transverse part of the self-energy tensor $\Sigma^{\mu\nu}$ of the $D_{s1}$ state, which is related to $\Sigma^{\mu\nu}$ through
\begin{align}
\Sigma^{\mu\nu}(k)=(g^{\mu\nu}-\frac{k^\mu k^\nu}{k^2})\Sigma^{T}(k^2)+\frac{k^\mu k^\nu}{k^2}\Sigma^{L}(k^2),
\end{align}
where $\Sigma^{L}(k^2)$ represents the longitudinal component of the self-energy.  Using the effective Lagrangians in Eq.~(\ref{eq1}),
the self-energy tensor $\Sigma^{\mu\nu}$ associated with the self-energy diagram shown in Fig.~\ref{cc2} is obtained as
\begin{align}
\Sigma^{\nu\sigma}_{D_{s1}}(k)&=-g^2_{D_{s1}}\epsilon_{\mu\nu\alpha\beta}\epsilon_{\eta\sigma\tau\delta}\int\frac{d^4q}{(2\pi)^4}\Phi^2[(q-k\omega_{D})^2_E]\nonumber\\
                      &\times{}k^{\mu}k^{\eta}(k-q)^{\alpha}(k-q)^{\tau}\frac{1}{q^2-m^2_{D}}\nonumber\\
                      &\times\frac{-g^{\beta\delta}+(k-q)^{\beta}(k-q)^{\delta}/m^2_{K^{*}}}{(k-q)^2-m^2_{K^{*}}}.
\end{align}
After a straightforward calculation, we obtain
\begin{align}
\Sigma^{T}(k^2)&=g^2_{D_{s1}}\int_{0}^{\infty}d\alpha\int_0^{\infty}d\beta\frac{1}{16\pi^2 z^2}\exp\{-\frac{1}{\Lambda^2}\nonumber\\
&\times{}[-2\omega_{D^0}^{2}k^2+\beta(m_{K^{*+}}^2-k^2)+\alpha m_{D^0}^2 +\frac{\Delta^{2}k^2}{4z}]\}\nonumber\\
&\times(-\frac{k^2\Lambda^2}{z}+\frac{k^4\Delta^2}{4z^2}-\frac{k^4\Delta^{2}}{4z^2})\label{eq6},
\end{align}
where $z=2+\alpha+\beta$ and $\Delta=-4\omega_{D}-2\beta$.

In addition to the interaction vertices described above, we also introduce the effective Lagrangian for the ${\cal{V}}{\cal{V}}{\cal{P}}$ and ${\cal{V}}{\cal{P}}{\cal{P}}$ couplings~\cite{Bando:1984ej,Bando:1987br,Nagahiro:2008cv}
\begin{align}
&\mathcal{L}_{{\cal{V}}{\cal{P}}{\cal{P}}} = -i g \langle {\cal{V}}_\mu [ {\cal{P}}, \partial^\mu {\cal{P}} ]\rangle\label{eq8},\\
&\mathcal{L}_{{\cal{V}}{\cal{V}}{\cal{P}}} = \frac{G^{'}}{\sqrt{2}}\epsilon^{\mu\nu\alpha\beta}\langle\partial_\mu {\cal{V}}_\nu \partial_\alpha {\cal{V}}_\beta {\cal{P}}\rangle,\\
&\mathcal{L}_{VVV} = i g \langle (V_\mu \overleftrightarrow{\partial}_\nu V^\mu ) V^\nu\rangle ,
\end{align}
where $\langle \cdots \rangle$ denotes the trace over the SU(4) flavor space. Here, the coupling constant $G^{'}$ is defined as $G^{'}=3g'^2/(4\pi^2f)$, with $g'=-G_Vm_\rho/(\sqrt{2}f^2)$~\cite{Nagahiro:2008cv},
where $G_V=55~\mathrm{MeV}$ and $f=93~\mathrm{MeV}$.  ${\cal{P}}$ and ${\cal{V}}$ denote the SU(4) pseudoscalar and vector meson fields, respectively. Their explicit forms are given by
\begin{equation}
P =\begin{pmatrix}
\frac{\pi^0}{\sqrt{2}} + \frac{\eta}{\sqrt{3}} + \frac{\eta'}{\sqrt{6}} & \pi^+ & K^+ & \bar{D}^0 \\
\pi^- & -\frac{\pi^0}{\sqrt{2}} + \frac{\eta}{\sqrt{3}} + \frac{\eta'}{\sqrt{6}} & K^0 & D^- \\
K^- & \bar{K}^0 & -\frac{1}{\sqrt{3}}\eta + \frac{2\eta'}{\sqrt{6}} & D_s^- \\
D^0 & D^+ & D_s^+ & \eta_c
\end{pmatrix},
\end{equation}
and
\begin{equation}
V^\mu =
\begin{pmatrix}
\frac{1}{\sqrt{2}}(\rho^0 + \omega) & \rho^+ & K^{*+} & \bar{D}^{*0} \\
\rho^- & \frac{1}{\sqrt{2}}(-\rho^0 + \omega) & K^{*0} & D^{*-} \\
K^{*-} & \bar{K}^{*0} & \phi & D_s^{*-} \\
D^{*0} & D^{*+} & D_s^{*+} & J/\psi
\end{pmatrix}^\mu .
\end{equation}
The coupling $g$ is fixed from the strong decay width of $K^* \to K\pi$.  With the help of Eq.~\ref{eq8}, the two-body decay width $K^{*+} \to K^0 \pi^+$ is related to $g$ as
\begin{equation}
\Gamma(K^{*+} \to K^0 \pi^+) = \frac{g^2}{6\pi m_{K^{*+}}^2}\, P_{\pi K^*}^3 = \frac{2}{3}\,\Gamma_{K^{*+}},
\end{equation}
where $P_{\pi K^*}$ is the three-momentum of the pion in the rest frame of the $K^*$ meson. Using the experimental value of the total decay width $\Gamma_{K^{*+}} = 50.3 \pm 0.8~\mathrm{MeV}$
and the hadron masses~\cite{ParticleDataGroup:2024cfk}, we obtain $g = 4.64$.

By combining all the components, the full decay amplitudes can be obtained straightforwardly. We first present the decay amplitude of the $D_{s1}$ as follows
\begin{align}
{\cal{M}}^{\pi^0}(&D^{*0}K^{+})=\frac{g^2g_{D_{s1}}}{2}\int\frac{d^4q}{(2\pi)^4}\Phi[(k_1\omega_{K^{*+}}-k_2\omega_{D^0})_E^2]\nonumber\\
                   &\times{}(k_1+q)^{\mu}\epsilon^{*}_{\mu}(p_1)\frac{i}{k_1^2-m^2_{D^0}}\epsilon_{\alpha\beta\eta\tau}p^{\alpha}\epsilon^{\beta}(p)k_2^{\eta}\nonumber\\
                   &\times{}\frac{i(-g^{\tau\lambda}+k_2^{\tau}k_2^{\lambda}/m^2_{K^{*+}})}{k_2^2-m^2_{K^{*+}}}(q+p_2)_{\lambda}\frac{i}{q^2-m^2_{\pi^0}},\\
{\cal{M}}^{\eta}(&D^{*0}K^{+})=\frac{2g^2g_{D_{s1}}}{3}\int\frac{d^4q}{(2\pi)^4}\Phi[(k_1\omega_{K^{*+}}-k_2\omega_{D^0})_E^2]\nonumber\\
                   &\times{}(k_1+q)^{\mu}\epsilon^{*}_{\mu}(p_1)\frac{i}{k_1^2-m^2_{D^0}}\epsilon_{\alpha\beta\eta\tau}p^{\alpha}\epsilon^{\beta}(p)k_2^{\eta}\nonumber\\
                   &\times{}\frac{i(-g^{\tau\lambda}+k_2^{\tau}k_2^{\lambda}/m^2_{K^{*+}})}{k_2^2-m^2_{K^{*+}}}(q+p_2)_{\lambda}\frac{i}{q^2-m^2_{\eta}},\\
{\cal{M}}^{\eta^{'}}(&D^{*0}K^{+})=\frac{-g^2g_{D_{s1}}}{6}\int\frac{d^4q}{(2\pi)^4}\Phi[(k_1\omega_{K^{*+}}-k_2\omega_{D^0})_E^2]\nonumber\\
                   &\times{}(k_1+q)^{\mu}\epsilon^{*}_{\mu}(p_1)\frac{i}{k_1^2-m^2_{D^0}}\epsilon_{\alpha\beta\eta\tau}p^{\alpha}\epsilon^{\beta}(p)k_2^{\eta}\nonumber\\
                   &\times{}\frac{i(-g^{\tau\lambda}+k_2^{\tau}k_2^{\lambda}/m^2_{K^{*+}})}{k_2^2-m^2_{K^{*+}}}(q+p_2)_{\lambda}\frac{i}{q^2-m^2_{\eta^{'}}},\\
{\cal{M}}^{\rho^0/\omega}(&D^{0}K^{+})=\frac{-G^{'}gg_{D_{s1}}}{2\sqrt{2}}\int\frac{d^4q}{(2\pi)^4}\Phi[(k_1\omega_{K^{*+}}-k_2\omega_{D^0})_E^2]\nonumber\\
                    &\times{}(k_1+p_1)_{\mu}\epsilon_{\alpha\beta\eta\tau}p^{\alpha}\epsilon^{\beta}(p)k_2^{\eta}\frac{i(-g^{\tau\lambda}+k_2^{\tau}k_2^{\lambda}/m^2_{K^{*+}})}{k_2^2-m^2_{K^{*+}}}\nonumber\\
                    &\times{}\epsilon_{\delta\sigma\theta\lambda}q^{\delta}k_{2}^{\theta}\frac{i(-g^{\mu\sigma}+q^{\nu}q^{\sigma}/m^2_{\rho^{0}/\omega})}{q^2-m^2_{\rho^{0}/\omega}}\frac{i}{k_1^2-m^2_{D^0}},\\
{\cal{M}}^{K^{-}}(&D_s^{*+}\eta)=\frac{2g^2g_{D_{s1}}}{\sqrt{3}}\int\frac{d^4q}{(2\pi)^4}\Phi[(k_1\omega_{K^{*+}}-k_2\omega_{D^0})_E^2]\nonumber\\
                    &\times{}(k_1+q)_{\mu}\epsilon_{\alpha\beta\eta\tau}p^{\alpha}\epsilon^{\beta}(p)k_2^{\eta}\frac{i(-g^{\tau\lambda}+k_2^{\tau}k_2^{\lambda}/m^2_{K^{*+}})}{k_2^2-m^2_{K^{*+}}}\nonumber\\
                    &\times{}(q+p_2)_{\lambda}\frac{i}{q^2-m^2_{K^{-}}}\frac{i}{k_1^2-m^2_{D^0}}\epsilon^{*\mu}(p_1),\\
{\cal{M}}^{K^{*-}}(&D_s^{*+}\eta)=\frac{G^{'2}g_{D_{s1}}}{2\sqrt{3}}\int\frac{d^4q}{(2\pi)^4}\Phi[(k_1\omega_{K^{*+}}-k_2\omega_{D^0})_E^2]\nonumber\\
                    &\times{}\epsilon_{\mu\nu\alpha\beta}q^{\mu}p_1^{\alpha}\epsilon^{*\beta}(p_1)\frac{i}{k_1^2-m^2_{D^0}}\epsilon_{\eta\tau\sigma\lambda}p^{\eta}\epsilon^{\tau}(p)k_2^{\sigma}\nonumber\\
                    &\times{}\epsilon_{\gamma\chi\xi\varpi}(k_2^{\gamma}q^{\xi}g_{\chi\theta}g_{\varpi\varphi}-q^{\gamma}k_2^{\xi}g_{\chi\varphi}g_{\varpi\theta})\nonumber\\
                    &\times{}\frac{i(-g^{\lambda\theta}+k_2^{\lambda}k_2^{\theta}/m^2_{K^{*+}})}{k_2^2-m^2_{K^{*+}}}\frac{i(-g^{\nu\varphi}+q^{\nu}q^{\varphi}/m^2_{K^{*-}})}{q^2-m^2_{K^{*-}}},\\
{\cal{M}}^{K^{*-}}(&D_s^{+}\eta)=\frac{gG^{'}g_{D_{s1}}}{\sqrt{6}}\int\frac{d^4q}{(2\pi)^4}\Phi[(k_1\omega_{K^{*+}}-k_2\omega_{D^0})_E^2]\nonumber\\
                    &\times{}(k_1+p_1)_{\mu}\epsilon_{\alpha\beta\eta\tau}p^{\alpha}\epsilon^{\beta}(p)k_2^{\eta}\frac{i(-g^{\tau\lambda}+k_2^{\tau}k_2^{\lambda}/m^2_{K^{*+}})}{k_2^2-m^2_{K^{*+}}}\nonumber
\end{align}
\begin{align}
                     &\times{}\epsilon_{\theta\delta\xi\varpi}(k_2^{\theta}q^{\xi}g_{\delta\lambda}g_{\varpi\nu}-k_2^{\xi}q^{\theta}g_{\varpi\lambda}g_{\delta\nu})\nonumber\\
                    &\times\frac{i(-g^{\mu\nu}+q^{\mu}q^{\nu}/m^2_{K^{*-}})}{q^2-m^2_{K^{*-}}}\frac{i}{k_1^2-m^2_{D^0}},\\
{\cal{M}}^{D_s^{*-}}(&D^{0}K^{+})=-\frac{gG^{'}g_{D_{s1}}}{2}\int\frac{d^4q}{(2\pi)^4}\Phi[(k_1\omega_{K^{*+}}-k_2\omega_{D^0})_E^2]\nonumber\\
                    &\times{}(k_1+p_2)_{\mu}\epsilon_{\alpha\beta\eta\tau}p^{\alpha}\epsilon^{\beta}(p)k_2^{\eta}\frac{i(-g^{\tau\lambda}+k_2^{\tau}k_2^{\lambda}/m^2_{K^{*+}})}{k_2^2-m^2_{K^{*+}}}\nonumber\\
                    &\times{}\epsilon_{\delta\lambda\sigma\theta}k_2^{\delta}q^{\sigma}\frac{i(-g^{\mu\theta}+q^{\mu}q^{\theta}/m^2_{D_s^{*-}})}{q^2-m^2_{D_s^{*-}}}\frac{i}{k_1^2-m^2_{D^0}},\\
{\cal{M}}^{D_s^{*-}}(&D^{*0}K^{+})=-g^2g_{D_{s1}}\int\frac{d^4q}{(2\pi)^4}\Phi[(k_1\omega_{K^{*+}}-k_2\omega_{D^0})_E^2]\nonumber\\
                    &\times{}[(p_1+q)_{\beta}g_{\alpha\eta}-(p_1+k_2)_{\alpha}g_{\beta\eta}+(k_2-q)_{\eta}g_{\alpha\beta}]\epsilon^{*}_{\eta}(p_1)\nonumber\\
                    &\times{}\frac{i(-g^{\beta\lambda}+k_2^{\beta}k_2^{\lambda}/m^2_{K^{*+}})}{k_2^2-m^2_{K^{*+}}}\epsilon_{\theta\tau\sigma\lambda}p^{\theta}\epsilon^{\tau}(p)k_2^{\sigma}(p_2+k_1)^{\chi}\nonumber\\
                    &\times{}\frac{i(-g^{\chi\alpha}+q^{\chi}q^{\alpha}/m^2_{D_s^{*-}})}{q^2-m^2_{D_s^{*-}}}\frac{i}{k_1^2-m^2_{D^0}},\\
{\cal{M}}(&D^{0}\pi^0K^{+})=\frac{gg_{D_{s1}}}{\sqrt{2}}\Phi[(k_1\omega_{K^{*+}}-k_2\omega_{D^0})_E^2]\epsilon_{\alpha\beta\eta\tau}p^{\alpha}\nonumber\\
                    &\times{}\epsilon^{\beta}(p)k_2^{\eta}\frac{i(-g^{\tau\lambda}+k_2^{\tau}k_2^{\lambda}/m^2_{K^{*+}})}{k_2^2-m^2_{K^{*+}}+im_{K^{*+}}\Gamma_{K^{*+}}}(p_1-p_2)_{\lambda},
\end{align}
\begin{align}
{\cal{M}}^{\rho^0/\omega}(&D^{*0}K^{+})=\frac{G^{'2}g_{D_{s1}}}{4}\int\frac{d^4q}{(2\pi)^4}\Phi[(k_1\omega_{K^{*+}}-k_2\omega_{D^0})_E^2]\nonumber\\
                            &\times{}\epsilon_{\mu\nu\alpha\beta}q^{\mu}p_1^{\alpha}\epsilon^{*\beta}(p_1)\epsilon_{\eta\tau\lambda\sigma}p^{\eta}\epsilon^{\tau}(p)k_2^{\lambda}\frac{i(-g^{\sigma\varpi}+k_2^{\sigma}k_2^{\varpi}/m^2_{K^{*+}})}{k_2^2-m^2_{K^{*+}}}\nonumber\\
                            &\times{}\epsilon_{\xi\kappa\chi\varpi}q^{\xi}k_{2}^{\chi}\frac{i(-g^{\nu\kappa}+q^{\nu}q^{\kappa}/m^2_{\rho^0/\omega})}{q^2-m^2_{\rho^0/\omega}}\frac{i}{k_1^2-m^2_{D^0}},\\
{\cal{M}}^{D^{*0}}(&D_s^{+}\eta)=\frac{G^{'2}g_{D_{s1}}}{\sqrt{6}}\int\frac{d^4q}{(2\pi)^4}\Phi[(k_1\omega_{K^{*+}}-k_2\omega_{D^0})_E^2]\nonumber\\
                            &\times{}\epsilon_{\mu\nu\alpha\beta}q^{\mu}k_2^{\alpha}\frac{i(-g^{\beta\lambda}+k_2^{\beta}k_2^{\lambda}/m^2_{K^{*+}})}{k_2^2-m^2_{K^{*+}}}\epsilon_{\eta\tau\sigma\lambda}p^{\eta}\epsilon^{\tau}(p)k_{2}^{\sigma}\nonumber\\
                            &\times{}\frac{i(-g^{\delta\nu}+q^{\nu}q^{\delta}/m^2_{D^{*0}})}{q^2-m^2_{D^{*0}}}(k_1+p_2)^{\delta}\frac{i}{k_1^2-m^2_{D^0}},\\
{\cal{M}}^{D^{*0}}(&D_s^{*+}\eta)=\frac{-g^2g_{D_{s1}}}{\sqrt{3}}\int\frac{d^4q}{(2\pi)^4}\Phi[(k_1\omega_{K^{*+}}-k_2\omega_{D^0})_E^2]\nonumber\\
                    &\times{}[(p_1+k_2)_{\beta}g_{\alpha\eta}-(p_1+q)_{\alpha}g_{\beta\eta}-(k_2-q)_{\eta}g_{\alpha\beta}]\epsilon^{*}_{\eta}(p_1)\nonumber\\
                    &\times{}\frac{i(-g^{\alpha\lambda}+k_2^{\alpha}k_2^{\lambda}/m^2_{K^{*+}})}{k_2^2-m^2_{K^{*+}}}\epsilon_{\tau\sigma\delta\lambda}p^{\tau}\epsilon^{\sigma}(p)k_2^{\delta}(p_2+k_1)^{\varphi}\nonumber\\
                    &\times{}\frac{i(-g^{\beta\varphi}+q^{\beta}q^{\varphi}/m^2_{D^{*0}})}{q^2-m^2_{D^{*0}}}\frac{i}{k_1^2-m^2_{D^0}},
\end{align}
where $f$ denotes the final state, while the superscript $i$ labels the exchanged particle contributing to the amplitude ${\cal{M}}^{i}(f)$. The four-momenta of the initial $D_{s1}$ state and its molecular constituents $D$ and $K^{*}$ are denoted by $p$, $k_1$, and $k_2$, respectively.  The final-state charmed meson and light meson carry momenta $p_1$ and $p_2$, while the exchanged particle in the $t$-channel has momentum $q$. For the three-body decay depicted in Fig.~\ref{cc1}(e), the momenta of the final-state $\pi$ and $K$ mesons are also assigned as $p_1$ and $p_2$, respectively.

Once the amplitudes are calculated, the partial decay widths for the two- and three-body decay processes can be evaluated using the following formulas:
\begin{align}
d\Gamma(D_{s1} &\to D^{(*)}K,D^{(*)}_{s}\eta) =\frac{1}{2J+1}\frac{1}{32\pi^2}\frac{|\boldsymbol{p}_1|}{m_{D_{s1}}^{2}}|\overline{\mathcal{M}}|^{2}d\Omega ,\\
d\Gamma(D_{s1}&\to D\pi{}K) =\frac{1}{2J+1}\frac{1}{(2\pi)^5}\frac{1}{16m^2_{D_{s1}}}\overline{|\mathcal{M}|^2}|\boldsymbol{p}_2^*||\boldsymbol{p}|\nonumber\\
                  &\times\,dm_{\pi{}K}\,d\Omega_{p_2}^*d\Omega_{D^0} ,
\end{align}
where $J$ is the total angular momentum of the $D_{s1}$, and
$|\boldsymbol{p}_1|$ is the three-momentum of the decay products in the center-of-mass frame.  The overline denotes the sum over the polarization vectors
of the final hadrons.  The quantities $(\boldsymbol{p}_2^*,\Omega_{p_2}^*)$ represent the momentum and angle of the particle $K$ in the rest frame of the
$\bar{K}^{*}$, respectively. The angle $\Omega_{D^0}$ denotes the angle of the $D^0$ meson in the rest frame of the decaying particle.  The variable $m_{\pi{}K}$
is the invariant mass of the $\pi K$ system, with the integration range $m_{\pi}+m_{K}\leq m_{12}\leq {m_{D_{s1}}}-m_{D^0}$ .

\section{RESULTS AND DISCUSSIONS}\label{Sec: results}
Within our theoretical framework, the cutoff parameter $\Lambda$ introduced in Eq.~\ref{eq2} is the only undetermined parameter that significantly affects the predictions.
Since it cannot be fixed from first-principles calculations, $\Lambda$ is constrained by fitting experimental observables. In this work, we determine $\Lambda$ by reproducing
the measured ratio
\[
R_{D_{s1}}=\frac{\Gamma(D_{s1}(2700)\to D^{*}K)}
{\Gamma(D_{s1}(2700)\to DK)}
=0.91\pm0.13\pm0.12 .
\]
The resulting $\Lambda$ is then used to calculate the total decay width of the $D_{s1}(2700)$, which is compared with the experimental value to examine whether the
$P$-wave $DK^{*}$ molecular scenario can describe its properties.

\begin{figure}[http]
\begin{center}
\includegraphics[bb=20 2 1050 350, clip, scale=0.35]{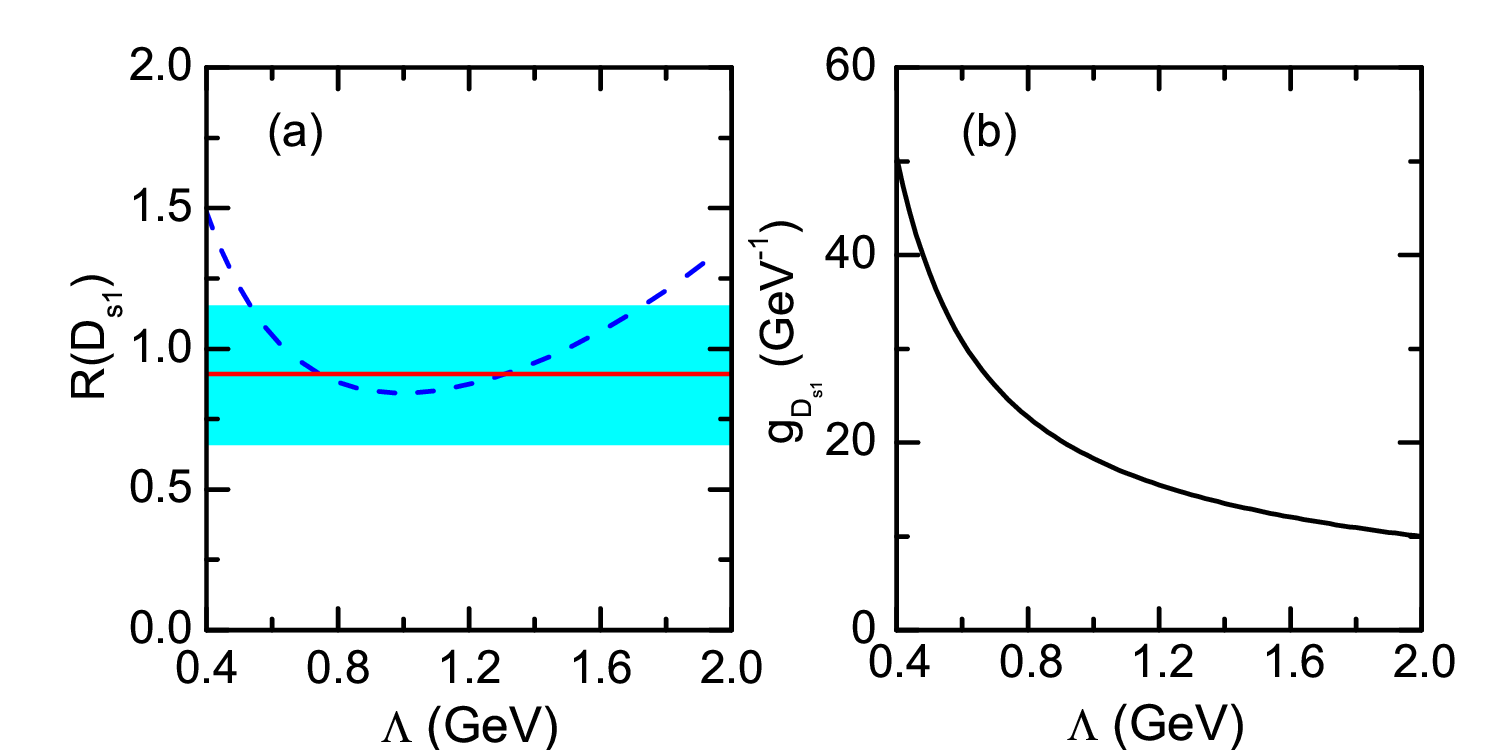}
\caption{(a) Theoretical prediction of $R_{D_{s1}}$ (blue dashed line) compared with the experimental measurement
(cyan band)~\cite{ParticleDataGroup:2024cfk}. The red solid line denotes the central value of the experimental result.
(b) The coupling constant $g_{D_{s1}}$ as a function of the cutoff parameter $\Lambda$.}\label{cc3}
\end{center}
\end{figure}

The theoretical result of $R_{D_{s1}}$ is presented in Fig.~\ref{cc3}(a). For comparison, the experimentally measured central value (red solid line) and its uncertainty (cyan band) are also shown. As illustrated in the figure, the calculated ratio
$R_{D_{s1}}=\Gamma(D_{s1}(2700)\to D^{*}K)/\Gamma(D_{s1}(2700)\to DK)$
exhibits a nonmonotonic dependence on the cutoff parameter $\Lambda$: it decreases initially and then increases with increasing $\Lambda$, reaching a minimum value of $R_{D_{s1}}=0.843$ at $\Lambda=1.00$ GeV, which is within the experimental uncertainty range of $0.66-1.16$. The experimental constraint is satisfied for $\Lambda=0.52-1.74$ GeV, and this cutoff range is adopted in the following calculations. Within this $\Lambda$ region, the calculated partial decay widths of the $DK$ and $D^{*}K$ channels vary from $3.936$ to $38.998$ MeV and from $4.641$ to $45.250$ MeV, respectively.

It should be noted that we explicitly calculate the partial decay widths of the channels $D_{s1}^{+}(2700)\to K^{+}D^{0}$ and $K^{+}D^{*0}$, while the corresponding charge partners, $K^{0}D^{+}$ and $K^{0}D^{*+}$, are obtained using isospin symmetry. Summing over all charge states gives the total decay widths $\Gamma(D_{s1}(2700)\to DK)$ and $\Gamma(D_{s1}(2700)\to D^{*}K)$, which enter the definition of $R_{D_{s1}}$. The same procedure is applied in the subsequent calculations.

Considering the range of $\Lambda$ values adopted in this work, we first examine the behavior of the calculated coupling constants. By substituting Eq.~\ref{eq6} into Eq.~\ref{eq3},
the dependence of the coupling constants on $\Lambda$ is obtained. After performing the integration over the parameters $\alpha$ and $\beta$ from 0 to infinity, the numerical results
of the coupling constants are shown in Fig.~3(b) for $\Lambda$ values ranging from $0.4$ to $2.0$ GeV. It is observed that the coupling constants decrease as $\Lambda$ increases and
display a strong dependence on the cutoff parameter.  More specifically, within the adopted range of $\Lambda=0.52-1.74$ GeV, the coupling constants decrease from $50.526$ at $\Lambda=0.52$
GeV to $11.226$ at $\Lambda=1.74$ GeV, demonstrating a pronounced dependence on the cutoff parameter.

\begin{figure}[http]
\begin{center}
\includegraphics[bb=-60 2 1050 395, clip, scale=0.33]{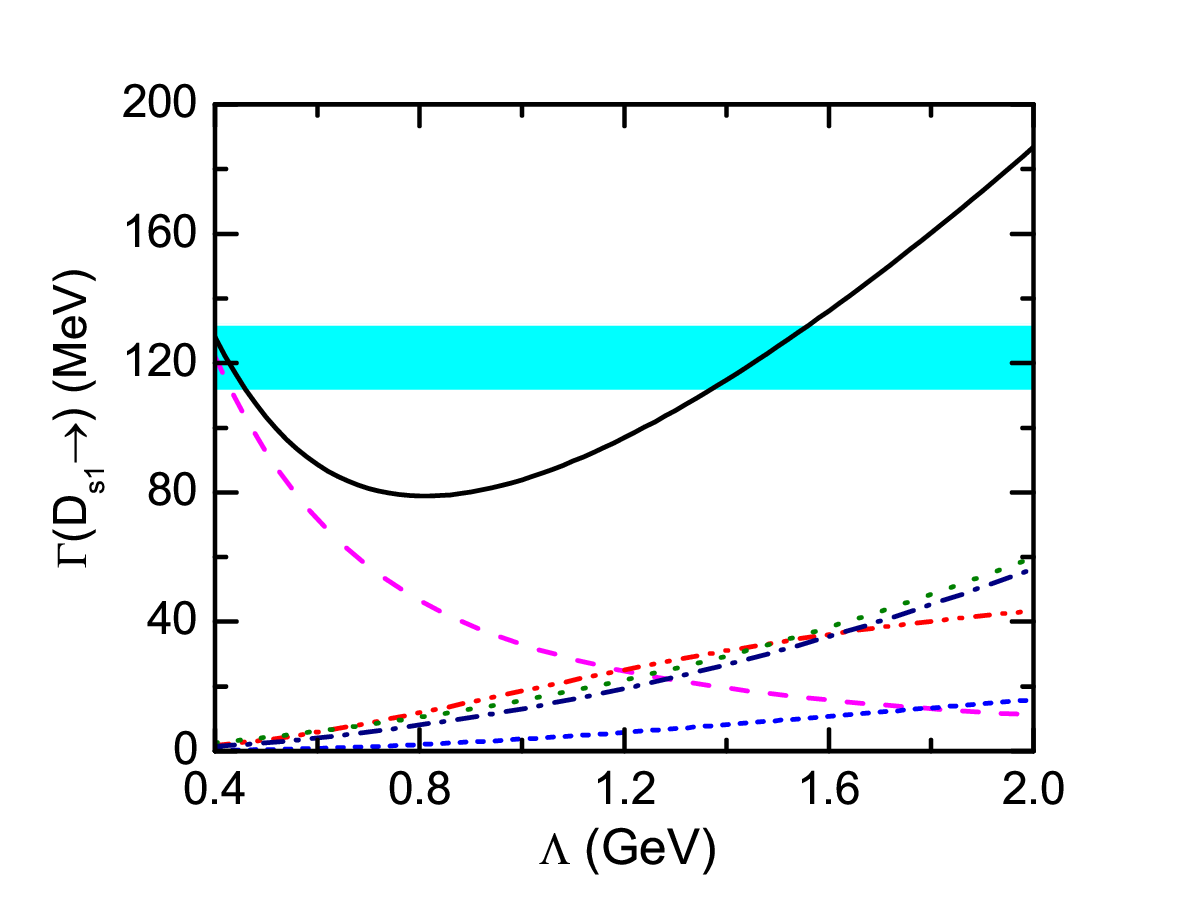}
\caption{Decay widths of $D_{s1}(2700)$ for the final states $DK$, $D^{*}K$, $D_{s}\eta$, $D^{*}_{s}\eta$, and $D\pi{}K$, as well as the total decay width. The cyan bands denote the corresponding
experimental values and their uncertainties~\cite{ParticleDataGroup:2024cfk}.}\label{cc4}
\end{center}
\end{figure}
After determining the coupling constants, we calculate the partial decay widths of the two-body decay channels $D_{s1}(2700)\to DK$ (Red dash dot dot line), $D^{*}K$ (Olive dot line),
$D_{s}\eta$ (Blue short dash line), and $D^{*}_{s}\eta$ (Navy dash dot line), as well as the three-body decay channels $D_{s1}(2700)\to D\pi{}K$ (Magenta dash line). The numerical results
are presented in Fig.~\ref{cc4} as functions of the parameter $\Lambda$ in the range of $0.52-1.74$ GeV.  The total decay width of $D_{s1}(2700)$ is obtained by summing all the partial decay
widths and is also shown in Fig.~\ref{cc4}.

From the results shown in Fig.~\ref{cc4}, it can be seen that the theoretical total decay width, represented by the solid black line, exhibits a nonmonotonic dependence on the cutoff
parameter $\Lambda$.  As $\Lambda$ increases, the total decay width initially decreases, reaching a minimum value, and subsequently increases.  Numerically, the decay width decreases
from $\Gamma(D_{s1}(2700))=99.673$ MeV at $\Lambda=0.52$ GeV to its minimum value of $\Gamma(D_{s1}(2700))=78.861$ MeV at $\Lambda=0.82$ GeV.  With further increasing $\Lambda$, the
decay width rises and reaches the lower boundary of the experimental uncertainty, $\Gamma(D_{s1}(2700))=112.842$ MeV, at $\Lambda=1.38$ GeV, and attains a maximum value of $152.758$ MeV
at $\Lambda=1.74$ GeV within the considered range.

The comparison between the theoretical predictions and the experimental measurements suggests that the $D_{s1}(2700)$ can be reasonably interpreted as a $P$-wave $DK^{*}$ molecular state.
In particular, within the cutoff range of $\Lambda=1.38-1.58$ GeV, the calculated total decay widths are in excellent agreement with the experimental value. Furthermore, the predicted ratio
$R_{D_{s1}}=\Gamma(D_{s1}(2700)\to D^{*}K)/\Gamma(D_{s1}(2700)\to DK)$ within this cutoff range is found to be in the range of $0.940-1.049$, which is in good agreement with the experimental
measurement of $0.91\pm0.13\pm0.12$. The simultaneous agreement of both the total decay width and the decay-width ratio with the experimental data lends further support to the possible
interpretation of the $D_{s1}(2700)$ as a $P$-wave $DK^{*}$ molecular state.

It is also worth noting that, as shown in Fig.~\ref{cc4}, the three-body decay width of the $D_{s1}(2700)\to D\pi K$ channel exhibits a strong dependence on the cutoff parameter $\Lambda$.
Specifically, it decreases from $87.732$ MeV at $\Lambda=0.52$ GeV to $13.791$ MeV at $\Lambda=1.74$ GeV, while the two-body decay widths show a mild increase with increasing $\Lambda$.
Consequently, for relatively small cutoff values ($\Lambda=0.52-1.20$ GeV), the $D\pi K$ channel becomes the dominant decay mode within the $DK^{*}$ molecular interpretation. However, this
decay mode has not yet been experimentally observed. As $\Lambda$ increases beyond $1.20$ GeV, the dominant contribution shifts to the experimentally observed two-body decay channels: the
$DK$ channel dominates for $\Lambda=1.20-1.52$ GeV, whereas the $D^{*}K$ channel becomes dominant for $\Lambda=1.52-1.74$ GeV.

Interestingly, although the $D_{s1}(2700)\to D_s^{*}\eta$ channel has the smallest phase space among the considered decay modes, its decay width is significantly larger than that of the
$D_{s1}(2700)\to D_s\eta$ channel. Specifically, within the cutoff range $\Lambda=0.52-1.74$ GeV, the predicted width of the $D_s^{*}\eta$ channel varies from $2.847$ to $42.167$ MeV,
which is comparable to those of the experimentally observed $DK$ and $D^{*}K$ channels. Nevertheless, this decay mode has not yet been experimentally explored. Therefore, the measurement
of the $D_s^{*}\eta$ final state may provide an important test of the $P$-wave $DK^{*}$ molecular interpretation of the $D_{s1}(2700)$.

\begin{figure}[http]
\begin{center}
\includegraphics[bb=-60 2 1050 395, clip, scale=0.35]{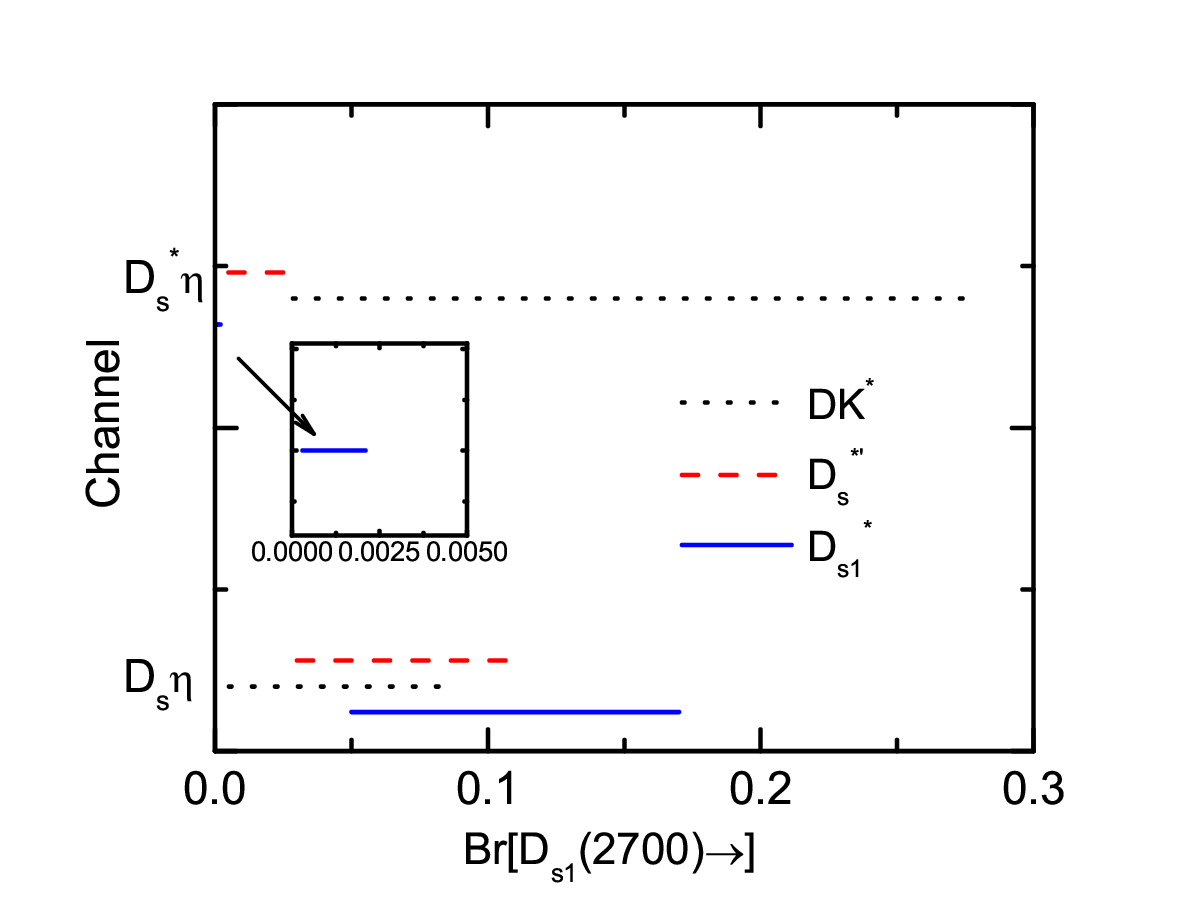}
\caption{Decay ratios of the $D_{s1}(2700)\to D_s^{(*)}\eta$ channels under the $P$-wave $DK^{*}$ molecular interpretation (black dotted line) and the conventional $(2S,J^P=1^-)$ quark-state
assignment~\cite{Colangelo:2007ds} (other curves). $D_{s}^{*'}$ and $D_{s1}^{*}$ denote the two heavy-quark symmetry partners associated with the light-quark angular momentum configurations
$s_l^P=1/2^-$ and $3/2^-$, respectively, within the $D_{s1}(2700)$ state.}\label{cc5}
\end{center}
\end{figure}
To further examine its discriminating power, we compare our prediction with two conventional quark-model descriptions~\cite{Colangelo:2007ds,Zhong:2009sk}, which provide good agreement
with the experimental $R_{D_{s1}}$ and total decay width.  We first consider the conventional $(2S,J^P=1^-)$ quark-state interpretation~\cite{Colangelo:2007ds}, which gives $R_{D_{s1}}=0.91\pm0.04$.
The comparison is shown in Fig.~\ref{cc5}. Our predicted $D_{s1}(2700)\to D_s^{*}\eta$ decay width differs significantly from the quark-model prediction, regardless of whether the light-quark
angular momentum configuration is $s_l^P=1/2^-$ or $3/2^-$. In contrast, the predicted $D_{s1}(2700)\to D_s\eta$ decay width shows no significant difference between the two interpretations,
and therefore this channel cannot serve as an effective discriminator.

Another conventional interpretation considers the $D_{s1}(2700)$ as a mixed state of the $2^3S_1$ and $1^3D_1$ configurations~\cite{Zhong:2009sk}. In this scenario, the calculated total
decay width and decay ratio,
\begin{align}
\Gamma[D_{s1}(2700)]\simeq133\pm22~\mathrm{MeV},\qquad
R_{D_{s1}}\simeq0.91\mp0.25,\nonumber
\end{align}
are also consistent with the experimental measurements~\cite{ParticleDataGroup:2024cfk}. However, the predicted $D_{s1}(2700)\to D_s^{*}\eta$ decay width is extremely small and nearly
vanishes, whereas the $D_{s1}(2700)\to D_s\eta$ decay width remains sizable. This behavior is qualitatively different from the prediction of the $P$-wave $DK^{*}$ molecular interpretation,
demonstrating the potential of the $D_s^{*}\eta$ channel as a discriminator of the internal structure of the $D_{s1}(2700)$.

\section{Summary}\label{sec:summary}
In this work, we investigate whether the $D_{s1}(2700)$ can be interpreted as a $DK^{*}$ molecular state by studying its strong decay properties. Under the $DK^{*}$ molecular assignment,
we calculate the partial decay widths of the $D_{s1}(2700)$ into the two-body final states $DK$, $D^{*}K$, $D_s\eta$, and $D_s^{*}\eta$, as well as the three-body final state $D\pi K$,
where some decay modes proceed through hadronic loop mechanisms. These decay processes are described by the $t$-channel exchanges of light mesons, including $\pi$, $\eta^{(\prime)}$, $K$,
$\rho$, $\omega$, and $K^{*}$, as well as heavy $D^{(*)}$ and $D_s^{(*)}$ mesons, together with tree-level contributions.

By comparing the calculated ratio $R_{D_{s1}}$ with the experimental measurement, we constrain the cutoff parameter to the range $\Lambda=0.52-1.74$ GeV. Within this range, the calculated
total decay width is consistent with the experimental result. In particular, for $\Lambda=1.38-1.58$ GeV, the theoretical predictions simultaneously reproduce both $R_{D_{s1}}$ and the total
decay width.  We further find that the decay patterns exhibit a strong dependence on the cutoff parameter. For relatively small cutoff values ($\Lambda\lesssim1.20$ GeV), the three-body decay
channel $D\pi K$ becomes the dominant decay mode, whereas with increasing $\Lambda$, the two-body channels $DK$ and $D^{*}K$ gradually become dominant.

Most importantly, the $D_s^{*}\eta$ channel, which has not yet been experimentally observed, is particularly noteworthy. Our calculations show that the predicted $D_{s1}(2700)\to D_s^{*}\eta$
decay width is not only comparable to those of the observed $DK$ and $D^{*}K$ channels, indicating its potential accessibility in future experiments, but also differs significantly from the
predictions of conventional quark-model interpretations, including both the pure $(2S,J^P=1^-)$ assignment and the mixed $2^3S_1-1^3D_1$ configuration.  Therefore, future measurements of the
$D_s^{*}\eta$ final state could provide a crucial experimental test for distinguishing between the molecular interpretation and conventional quark-state assignments of the $D_{s1}(2700)$.

In summary, our results support the possible interpretation of the $D_{s1}(2700)$ as a $P$-wave $DK^{*}$ molecular state. In particular, the $D_s^{*}\eta$ decay channel, which is highly
sensitive to the internal structure of the state, provides an important probe for revealing the nature of the $D_{s1}(2700)$.

\section*{Acknowledgments}
This work was supported by the Sailing Plan Project of Yibin University (No. 2021QH06). Y. Huang acknowledges support from the National Natural Science Foundation of China under Grant No. 12005177,
as well as support from the Fundamental Research Funds for the Central Universities under Grant No. 2682026TPY011.

%
\end{document}